\documentclass[14pt]{extarticle}
\usepackage[letterpaper, margin=1in]{geometry}
  
\usepackage{graphicx}
\usepackage{amsmath}
\usepackage[hidelinks=true]{hyperref}
\usepackage{cleveref}
\usepackage{parskip}
\usepackage[square,numbers,sort&compress]{natbib}
\usepackage[charter]{mathdesign}
\usepackage{microtype}

\usepackage{tikz}
\usetikzlibrary{trees}
\usetikzlibrary{bayesnet}

\usepackage{authblk} 

\usepackage{titlesec}
\titleformat{\section}{\normalsize\bfseries}{\thesection}{1em}{}
\titleformat{\subsection}{\normalsize\bfseries}{\thesubsection}{1em}{}
\titleformat{\subsubsection}{\normalsize\bfseries}{\thesubsubsection}{1em}{}
\usepackage[font=small]{caption}

\title{Towards Multilevel Modelling of Train Passing Events on the Staffordshire Bridge}

\author[1]{Lawrence A. Bull}
\author[1]{Chiho Jeon}
\author[1,4]{Mark Girolami}
\author[2,4]{Andrew Duncan}
\author[1]{Jennifer Schooling}
\author[3]{Miguel Bravo Haro}

\affil[1]{University of Cambridge, Department of Engineering, Cambridge, CB3 0FA, UK}
\affil[2]{Imperial College London, Department of Mathematics, London, SW7 2AZ, UK}
\affil[3]{City, University of London, Department of Engineering, London, EC1V 0HB, UK}
\affil[4]{The Alan Turing Institute, The British Library, 96 Euston Road, London, NW1 2DB, UK}

\date{May 2023}

\begin{document}

\maketitle

\begin{abstract}
We suggest a multilevel model, to represent aggregate train-passing events from the Staffordshire bridge monitoring system. %
We formulate a combined model from simple units, representing strain envelopes (of each train passing) for two types of commuter train. %
The measurements are treated as a longitudinal dataset and represented with a (low-rank approximation) hierarchical Gaussian process. %
For each unit in the combined model, we encode domain expertise as boundary condition constraints and work towards a general representation of the strain response. %
Looking forward, this should allow for the simulation of train types that were previously unobserved in the training data. %
For example, trains with more passengers or freights with a heavier payload. The strain event simulations are valuable since they can inform further experiments (including FEM calibration, fatigue analysis, or design) to test the bridge in hypothesised scenarios. %
\end{abstract}

\vfill
\textbf{Keywords:} Multilevel Models, Uncertainty Quantification, Bridge Monitoring, Simulation, Digital Twins, Meta-Analysis.
\vspace{4em}

\section{Introduction}

An increasing number of bridge-monitoring projects utilise streaming telemetry data \cite{butler2016integrated, butler2018monitoring, cross2012structural, dervilis2015robust}. %
It is vital that we develop appropriate statistical models to represent and extract valuable insights from these large datasets, since the bridges constitute critical infrastructure within modern transportation networks. %
The process of monitoring engineered systems via streaming data is typically referred to as Structural Health Monitoring (SHM) and while successful applications have been emerging in recent years, a number of challenges remain for practical implementation~\cite{sohn2003review}. %
During model design, these concerns usually centre around \textit{low variance} data: that is, measurements are not available for the entire range of expected operational, environmental, and damage conditions. %
Consider a bridge following construction, this will have a relatively small dataset that should only be associated with \textit{normal} operation. %
On the other hand, a structure with historical data might still not experience low-probability events -- such as extreme weather or landslides.

An obvious solution considers sharing data (or information) between structures; this has been the focus of a large body of recent work \cite{pt1,pt2,pt3}. %
Since no two structures are identical, simply combining the data (complete pooling) is rarely sufficient\footnote{Although, in certain cases, complete pooling works, especially for more simple operations such as novelty detection with consistent data \cite{pt1,bull2022bayesian}}. %
A number of statistical and ML approaches become suitable in this setting: domain adaptation \cite{pt3}, transfer learning \cite{tsialiamanis2020partitioning}, multitask learning \cite{bull2023hierarchical}, and federated learning \cite{anaissi2023personalised} are a few, recently explored examples. %

In this article, we present a model design for data from the landmark IB5 (Staffordshire) rail bridge digital twin \cite{butler2018monitoring,lin2019performance}. %
A convenient artefact of rail bridge monitoring is the repeated \textit{and comparable} train passing events, which readily allow information to be shared in ways that are less feasible for foot or automobile bridges. %
Train events occur frequently, with an average of 60-70 trains crossing IB5 per day. %
Many of these correspond to passenger/commuter trains, the focus of this initial work. %
We propose a Gaussian process representation, to work towards encoding prior knowledge of train events within a general, flexible representation. %

The layout of this paper is as follows. \Cref{s:IB5} outlines the existing IB5 sensing system; \Cref{s:classify} summarises a decision tree to classify train events; \Cref{s:modelling} formulates the multilevel Gaussian process representation; \Cref{s:results} presents initial monitoring insights from the model; \Cref{s:conc} offers concluding remarks and summarises future work. %

\section{The Norton (IB5) Bridge}\label{s:IB5}

It is widely accepted that monitoring and analysing telemetry data from bridges provides valuable insights. %
Though in practice, there are few examples of permanent and reliable sensing systems, which allow for the remote acquisition of \textit{streaming data}. %
The Norton intersection bridge 5 (IB5) is one example, with a sensing system that continuously monitors signals, such that train events are recorded, stored, and made accessible via an API \cite{broo2022design}. %

\begin{figure}[h]
    \centering\includegraphics[width=.8\linewidth]{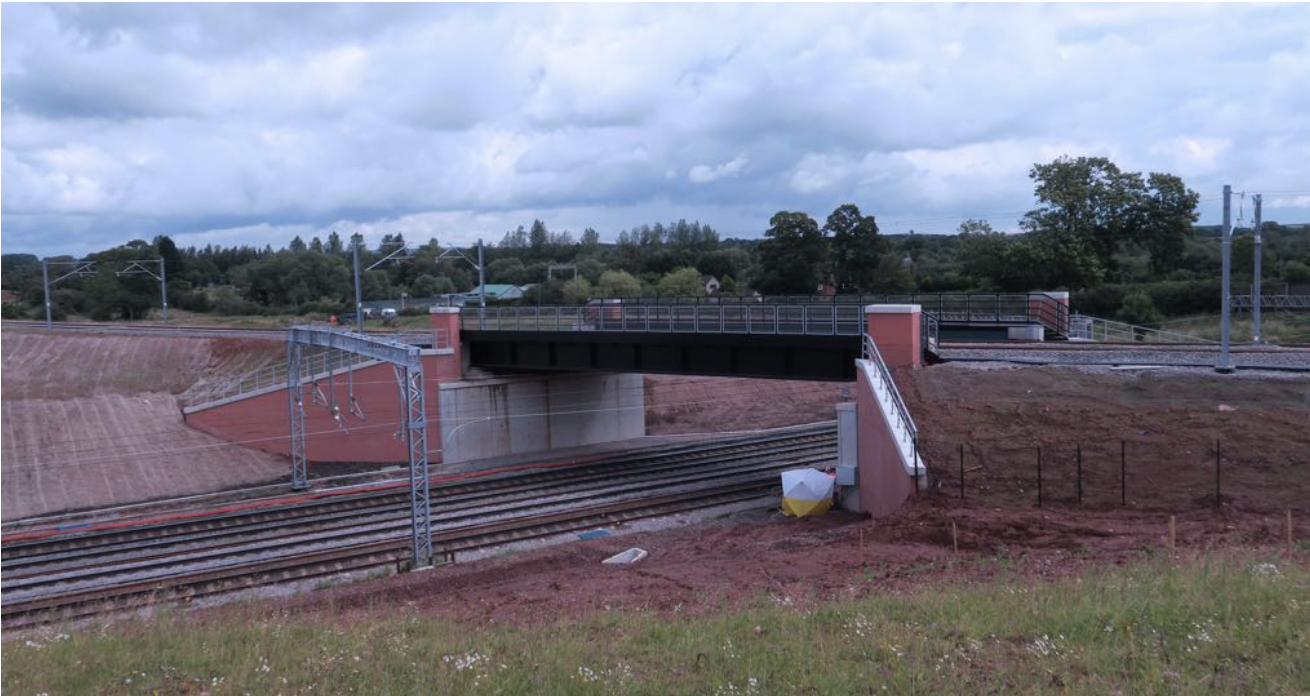}
    \caption{The IB5 Norton bridge, a composite structure with steel I-girders and a \textit{cast-in-situ} reinforced concrete deck, spanning 26.8m}
    \label{fig:ib5IRL}
\end{figure}

The IB5 structure itself is a half-through, E-type bridge owned by Network Rail, with a single skew axis span of 26.8~m, shown in \Cref{fig:ib5IRL}. %
It carries two rail lines on the West Coast Main Line near Crewe, UK, constructed as part of a rail redevelopment project -- the \href{https://www.railway-technology.com/projects/stafford-area-improvement-programme-saip/}{Stafford Area Improvement Programme}. %

The bridge has been instrumented with numerous sensors. %
Fibre Bragg Gratings (FBGs) were installed during construction. %
These were attached to the main and cross beams, and embedded in the reinforced concrete deck. %
More recently, accelerometers were deployed throughout the main load-bearing steel girder, as well as optical, humidity, and temperature sensors (and video cameras) \cite{fidler2022augmenting}. %

The acquisition system monitors all sensors in real-time, capturing a synchronised batch of raw measurements with each train passing. %
Over 34,000 passing events have been captured at the time of writing. %

\subsection{Strain and BWIM Data}

This work considers strain data and features from a Bridge Weigh in Motion (BWIM) system, implemented in previous phases of the IB5 project~\cite{huseynov2023setting}. %
FBG raw measurements are \textit{changes in wavelength} $\Delta \lambda$, which we consider from the west main girder of the bridge only. %
This is converted to the \textit{change in micro-strain} ($\Delta \mu \varepsilon$) as follows~\cite{lin2019performance},
\begin{align}
    \Delta \mu \varepsilon =\frac{1}{k_{\varepsilon}} \cdot \frac{\Delta \lambda}{\lambda_0} \times 10^{-6}
\end{align}
where $k_\epsilon$ is the gauge factor provided by the FBG manufacturer (0.78), and $\lambda_0$ scales raw measurements ($\Delta \lambda$) to present a relative change. %
Since we hope to compare the bridge response for many events, the reference $\lambda_0$ is an average from the start of each time series. %
In practical terms, this scales each response to zero at the start of the strain envelope. %
Also, note that FBG sensors are highly sensitive to temperature variations; however, here we assume that changes in temperature are negligible for each event, and leave out the associated compensation (which can be found in~\cite{lin2019performance}). %

\begin{figure}[ht]
    \centering
    \includegraphics[width=.9\linewidth]{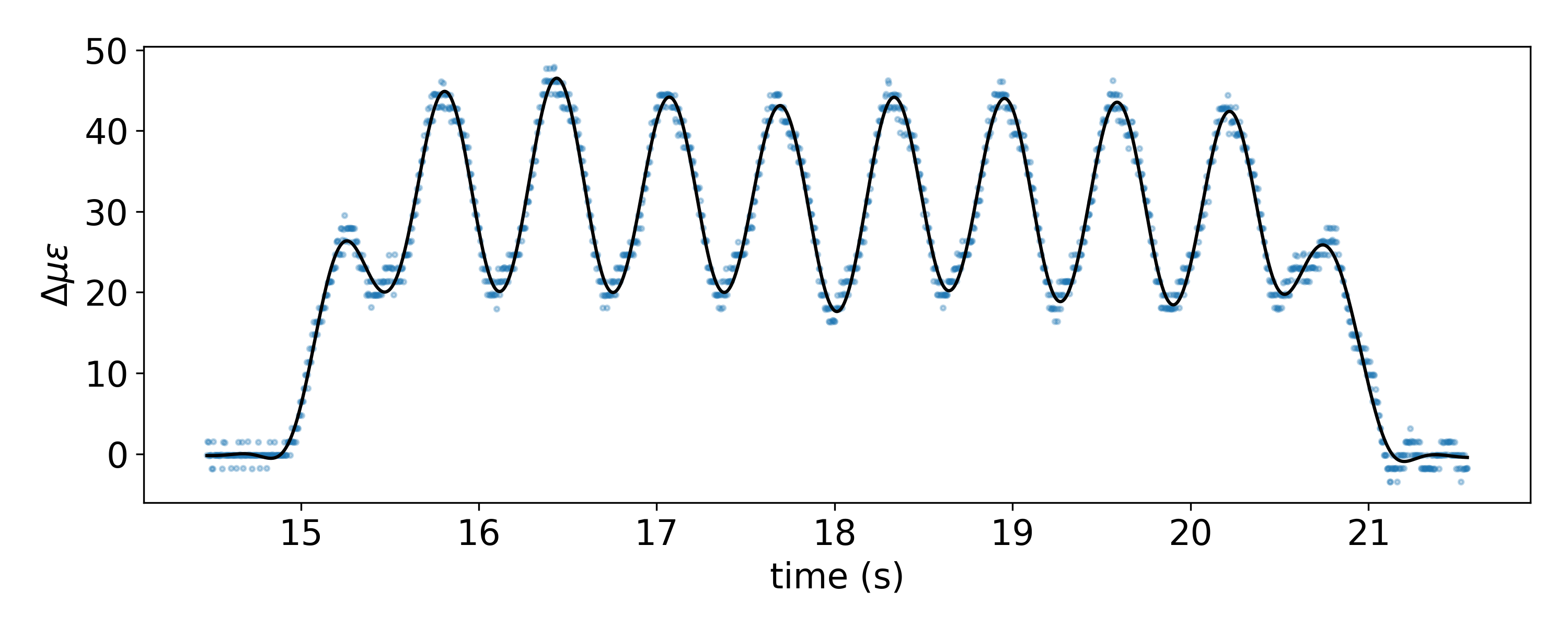}
    \caption{Strain time-series data (blue) and the filtered signal (black).}
    \label{fig:time-strain-eg}
\end{figure}

\Cref{fig:time-strain-eg} presents an example of the strain data over time. %
The underlying function has a convenient interpretation: the sum of \textit{strain influence lines} per axle, which can be used to estimate each axle weight contribution \cite{obrien2006calculating}. %
Currently, strain influence lines are only utilised within the BWIM system, to classify passing events according to train type (\Cref{s:classify}). 
However, looking forward, parametrising the function as a sum over (influence line) bases presents an interesting modelling perspective -- discussed as future work in \Cref{s:modelling,s:conc}. %

Conceptually, the BWIM system turns the bridge into a weighing scale, estimating each axle weight from \textit{in-situ} monitoring data. %
The result is a number of insightful features, three are considered in this work:
(i) number of axles,
(ii) axle weights,
(iii) axle spacing,
and transformations thereof. %

\section{Train Event Classification}\label{s:classify}

\begin{figure}[ht]
    \resizebox{\linewidth}{!}{%
    \begin{tikzpicture}[
    grow'=right,
    edge from parent fork right,
    level distance=7.5em,
    level 1/.style={sibling distance=4em},
    level 2/.style={sibling distance=4em},
    every node/.style={rectangle, draw, align=center}
    ]
    \node {16 axles?}
        child {node {yes} 
            child {node {axl-sep $<$ 5.3}
                child {node {yes}
                    child {node {\color{teal} 350 trains}}}
                child {node {no}
                    child {node {\color{purple} 22x trains}}}
          }}
        child {node {no}
          child {node {\textit{other} trains}}
        };
    \end{tikzpicture}}
    \caption{Train classification using BWIM data: [axl-sep] is the average axle separation, and 22x refers to either 220 or 221 trains.}
    \label{fig:decision-tree}
\end{figure}
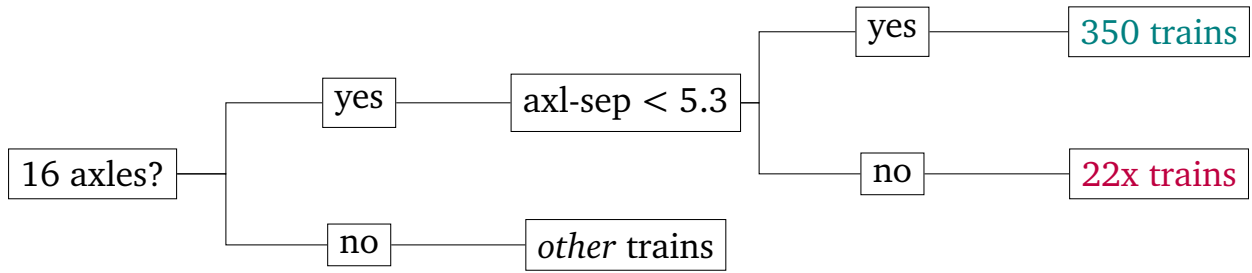

The focus of this article is to design a general representation of train events. %
With this in mind, it makes sense to initially consider \textit{typical} strain responses, and then alter the representation, if required (ideally without increasing model complexity). %
An obvious starting point is 16-axle commuter train types; including 350, 220, and 221 trains~\cite{lin2019performance}. %
Given the BWIM features extracted from each train passing, alongside knowledge of train design, a simple decision tree can be used to isolate the commuter train data. %
The categorisation is presented in \Cref{fig:decision-tree}.

\begin{figure}[ht]
    \centering
    \includegraphics[width=.9\linewidth]{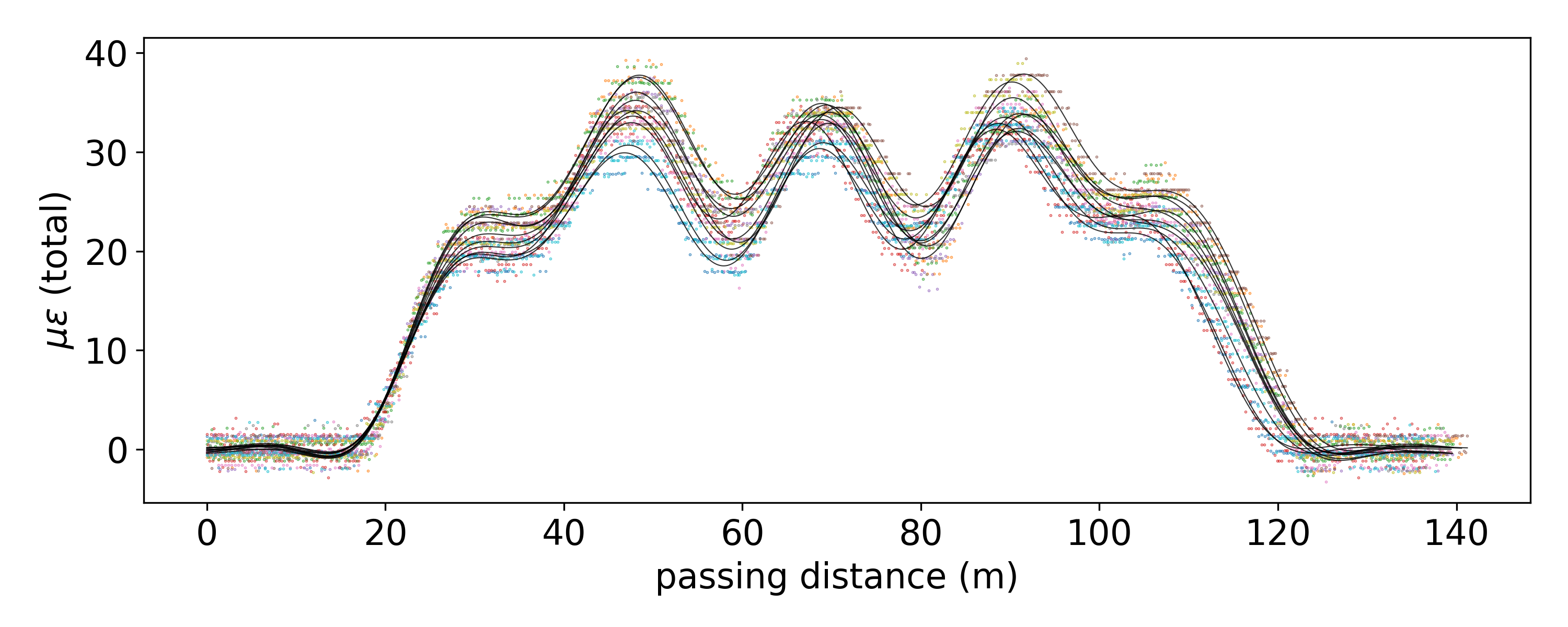}
    \caption{An example of ten passing events for 350 trains. Different colour markers for each train.}
    \label{fig:350trains}
\end{figure}

In the current work, we consider both 350 and 22x trains, highlighted with {\color{teal} green} and {\color{purple} purple} text at the leaves of the tree in \Cref{fig:decision-tree}. %
An example of ten 350 passing events is shown in \Cref{fig:350trains}. %
Note the input normalisation, using passing distance rather than time, which horizontally scales each strain signal to be comparable, irrespective of the train-speed. %

Observing \Cref{fig:350trains}, it becomes clear that the strain data are characteristic and repeatable (in this case, for 350 trains). %
In turn, their underlying function should be ideal for both monitoring and finding a general representation. %
Here we explore a basic Gaussian process representation, which captures the characteristic response, alongside an appropriate uncertainty quantification for the longitudinal data. %
Before inference, we z-score normalise the inputs and divide the response by its scale, as shown in \Cref{fig:350and22x}. %
This allows for stable inferences via Hamiltonian Monte Carlo and convenient prior definitions in the following section. %

\section{Model Design}\label{s:modelling}

We now describe the multilevel Gaussian process to represent the aggregated data. %
The multilevel structure will naturally extend to trains of the same type, while also presenting options to incorporate different designs. %
Importantly, with some (mean function) modifications, the model should represent the uncertainty of \textit{event-specific} functions, alongside the uncertainty due to variations between different events of the same train type, i.e.\ \textit{type-sepcific} functions. %

We use the notation $\mathbf{y}_k$ to denote the strain series (response vector) for event $k \in \{1,2,\ldots, K\}$ with the corresponding inputs $\mathbf{x}_k$ (passing distance). %
Each event is then approximated as some function $f_k$ with additive noise $\boldsymbol{\epsilon}_k$,
\begin{align}
    \mathbf{y}_k = f_k(\mathbf{x}_k) + \boldsymbol{\epsilon}_k
\end{align}

Following a standard Bayesian approach \cite{rasmussen2006gaussian} priors are placed over the latent functions and variables. %
Where the prior encodes our knowledge of the generating process \textit{before} the model is informed by data; then, by conditioning (or updating) the priors given measurements, we find the posterior distribution. %
The posterior can then be used to make predictions, which utilise a combination of (i) domain knowledge encoded in the prior and (ii) patterns in the observed data. %

First, a shared Gaussian process (GP) prior distribution is placed over each latent function,
\begin{align}
    f_k \sim \textrm{GP}\big(m(\cdot),\; k(\cdot, \cdot)\big)
\end{align}

This prior is fully specified by its mean $m(\cdot)$ and covariance functions $k(\cdot, \cdot)$. %
The mean captures the expected \textit{path} of the response, while the covariance describes the similarity between any two outputs, with respect to their input (i.e.\ the function's \textit{smoothness}). %
Domain expertise and knowledge of the underlying physics can (and should) be encoded within both of these functions; although, for now, we keep their definitions simple,
\begin{align}
    m(x_i) &= 0 \\
    k(x_i, x_j) &= \alpha^2\left(1+\frac{\sqrt{3}\left|\mathbf{x}_i-\mathbf{x}_j\right|}{l}\right) \exp \left(-\frac{\sqrt{3}\left|\mathbf{x}_i-\mathbf{x}_j\right|}{l}\right)\label{eq:Mat}
\end{align}

these correspond to a zero-mean and Mat\'{e}rn 3/2 kernel function. %
These prior functions have \textit{hyperparameters}: the \textit{process variance} $\alpha$ and \textit{length scale}, $l$ which have their own (hyper) priors,
\begin{align}
    \alpha \sim \textrm{Truncated-Normal}_+(1, 1), \qquad
    l \sim \textrm{Half-Normal}_+(1)
\end{align}
The prior distribution for the additive noise is also defined, and collected with other latent variables in $\boldsymbol{\theta}$,
\begin{align}
    \epsilon_{ki} \sim \textrm{Normal}(0, \sigma_k), \qquad
    &\sigma_k \sim \textrm{Half-Normal}_+(.2) \\[1em]
    \boldsymbol{\theta} \triangleq \big\{\alpha, l, \{\sigma_k&\}_{k=1}^K\big\}
\end{align}
Note, the half-normal distributions are zero-centred and the truncated normal has a minimum value of zero, since $\{\alpha, l, \{\sigma_k\}\}$ must be non-negative. %

Since the hyperparameters also have a Bayesian treatment, they are also updated given the training data. %
Practically, this means that the shape of the prior adapts during inference (via the mean and covariance functions). %
The above priors are set to be weakly informative, in view of the normalised space, postulating a signal-to-noise ratio of 5 (in terms of standard deviation). %

Importantly, $k$ subscripts are missing for $\{\alpha, l\}$, indicating that hyperparameters are \textit{shared} between all $k$ GPs (of both 350 and 22x events). %
A convenient interpretation is that the shared prior captures the characteristics of \textit{typical} commuter trains, which should be useful in monitoring procedures (especially novelty detection). %
Looking forward, $m(\cdot)$ should be specified beyond a zero-mean assumption (with parameters contributing to $\boldsymbol{\theta}$) to present a general mean function, which might represent a generic 350 train or 22x train specifically. %
The semi-parametric formulation \citet{kennedy2001bayesian} will be considered, alongside a basis function interpretation \cite{obrien2006calculating}. %


\subsection{A low-rank implementation to encode boundary conditions}

The number of observations per event varies quite widely, depending on the speed of the train (750-1500). %
Since Gaussian processes scale badly with large data, we implement a low-rank approximation. %
Following \citet{riutort2023practical} another basis function approximation is used for the \textit{kernel function}, which reduces the order of complexity from ${O}(N^3)$ to ${O}(N \cdot M + M)$, where $M$ is the number of basis functions. %

As the Mat\'{e}rn 3/2 kernel is a stationary function, it can be approximated using an eigendecomposition. %
\citet{solin2020hilbert} achieve this by interpreting $k(\cdot, \cdot)$ as the kernel of a pseudo-differential operator. %
The pseudo-differential operator is then approximated via Hilbert space methods on a compact subset $\Omega \subset \textrm{R}^D$, subjected to boundary conditions. %
A convenient side-effect is that the boundary conditions can be utilised to constrain the approximation given domain knowledge, see~\citet{jones2023constraining} for an engineering example. %

Considering the z-score normalised inputs in \Cref{fig:350and22x}, we know that most inputs should vary between approximately $[-2, 2]$ while returning to zero-strain at the boundaries of the envelope. %
This scaling naturally defines an interval in which the approximated GP is valid,
\begin{align}
    \Omega \in [-L, L] = [-3, 3]
\end{align}

Note the \textit{buffer} for $L$ (such that $L>2$) is to support stable inferences and ensure that test data do not fall outside the approximating window of $\Omega$. %

Within $\Omega$, the kernel is then be approximated as follows~\cite{riutort2023practical},
\begin{align}
k\left(x_i, x_j\right) \approx \sum_{m=1}^M S\left(\sqrt{\lambda_m}\right)\phi_m(x_i)\phi_m\left(x_j\right)
\end{align}
where $x_i, x_j \in \Omega$; $S$ is the spectral density of the kernel function; $\lambda_m$ are the eigenvalues and $\phi_m(x)$ are the eigenvectors of the Laplacian operator in the given domain~$\Omega$. %
In these experiments, we approximate with 40 basis functions ($M=40$).

For a univariate Mat\'{e}rn 3/2 function, $S(\cdot)$ is as follows \cite{solin2020hilbert},
\begin{align}
S(\boldsymbol{\omega})=\alpha \frac{4 \cdot 3^{3 / 2}}{l^3}\left(\frac{3}{l^2}+\boldsymbol{\omega}^{\top} \boldsymbol{\omega}\right)^{-2}
\end{align}

Then, following \cite{riutort2023practical}, the following eigenvalue problem is considered, with \textit{Dirichlet} boundary conditions respecting $\Omega$, %
\begin{align}
-\nabla^2 \phi_m(x) & =\lambda_m \phi_m(x), & & x \in \Omega \\
\phi_m(x) & =0, & & x \notin \Omega
\end{align}
The solution of which presents the following analytical eigenvalues and eigenvectors,
\begin{align}
\lambda_m & =\left(\frac{m \pi}{2 L}\right)^2 \\
\phi_m(x) & =\sqrt{\frac{1}{L}} \sin \left(\sqrt{\lambda_m}(x+L)\right)
\end{align}
Practically, Dirichlet conditions are justified, since the strain envelopes are bounded to zero before and after each signal (in view of the normalisation and strain-conversion). %
Further details of the equivalent linear implementation can be found in \cite{riutort2023practical}. %


\section{Results: Towards Monitoring Procedures}\label{s:results}

\begin{figure}[ht]
    \centering
    \includegraphics[width=\linewidth]{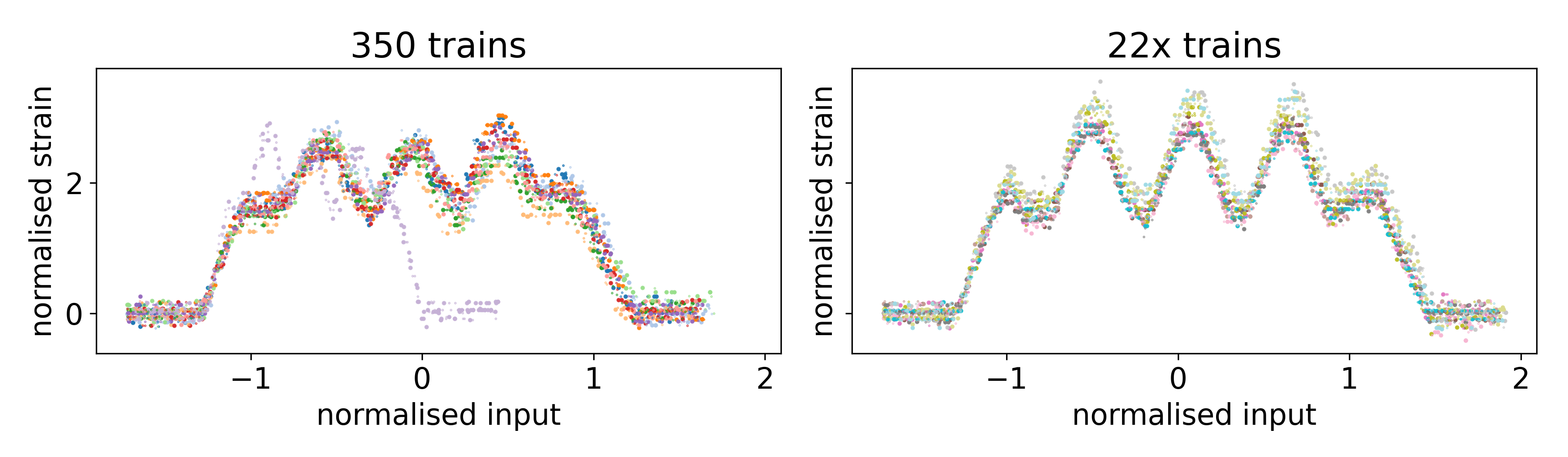}
    \caption{Ten 350-type and ten 22x-type train passing events: all modelled by the multilevel GP regression. Events are plotted with different colours. Pink markers in the left-hand plot correspond to an outlying event.}
    \label{fig:350and22x}
\end{figure}

The dataset comprises 20 passing events: ten 350 trains and ten 22x trains, sampled at random from the historical data and plotted in \Cref{fig:350and22x}. %
We intentionally include an outlying event, shown by the pink markers in the left panel. %
These outlying data appear to relate to an erroneous speed measurement, and therefore, an incorrect scaling of the inputs. %
This outlier (within the 350 trains) is used to demonstrate simple monitoring capabilities. %

\begin{figure}
    \centering
    \includegraphics[width=\linewidth]{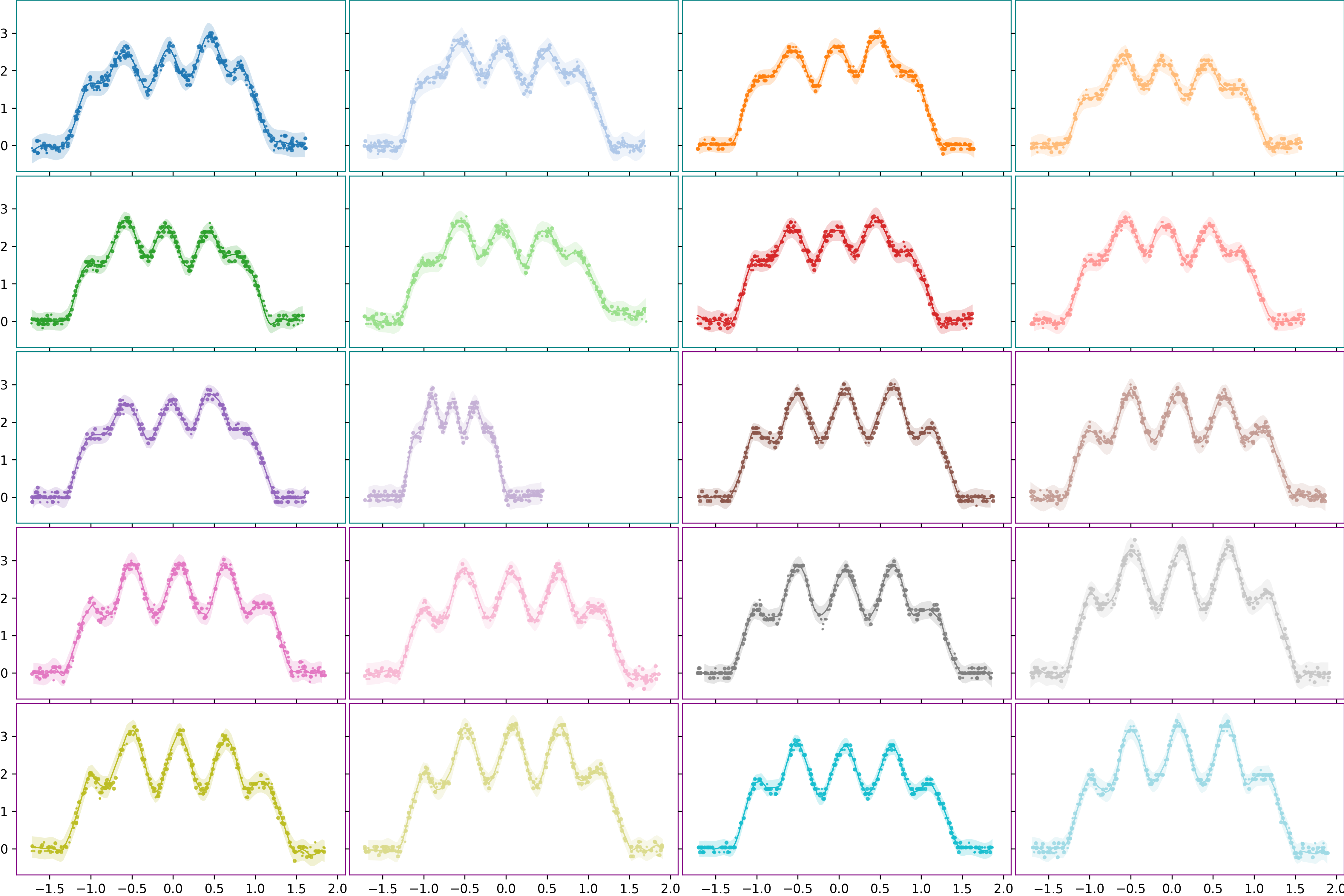}
    \caption{Posterior predictive distributions for the multilevel GP: teal frames are for the 350 train types and purple frames are for 22x trains.}
    \label{fig:post_pred}
\end{figure}

The models were written in probabilistic programming language \texttt{Stan}~\cite{carpenter2017stan} and inferred via MCMC with the no U-turn implementation of Hamiltonian Monte Carlo~\cite{hoffman2014no}. %
Throughout, 1000 iterations are used for burn-in and inference. %
The posterior predictive distribution of each GP is shown in \Cref{fig:post_pred}. %
The plots indicate a good fit to the data, however, the representation at the start/end of each strain envelope is quite poor, since the GP must represent the entire function without a parametrised mean -- as discussed, this will be the focus of future work.

\begin{figure}
    \centering
    \includegraphics[width=.65\linewidth]{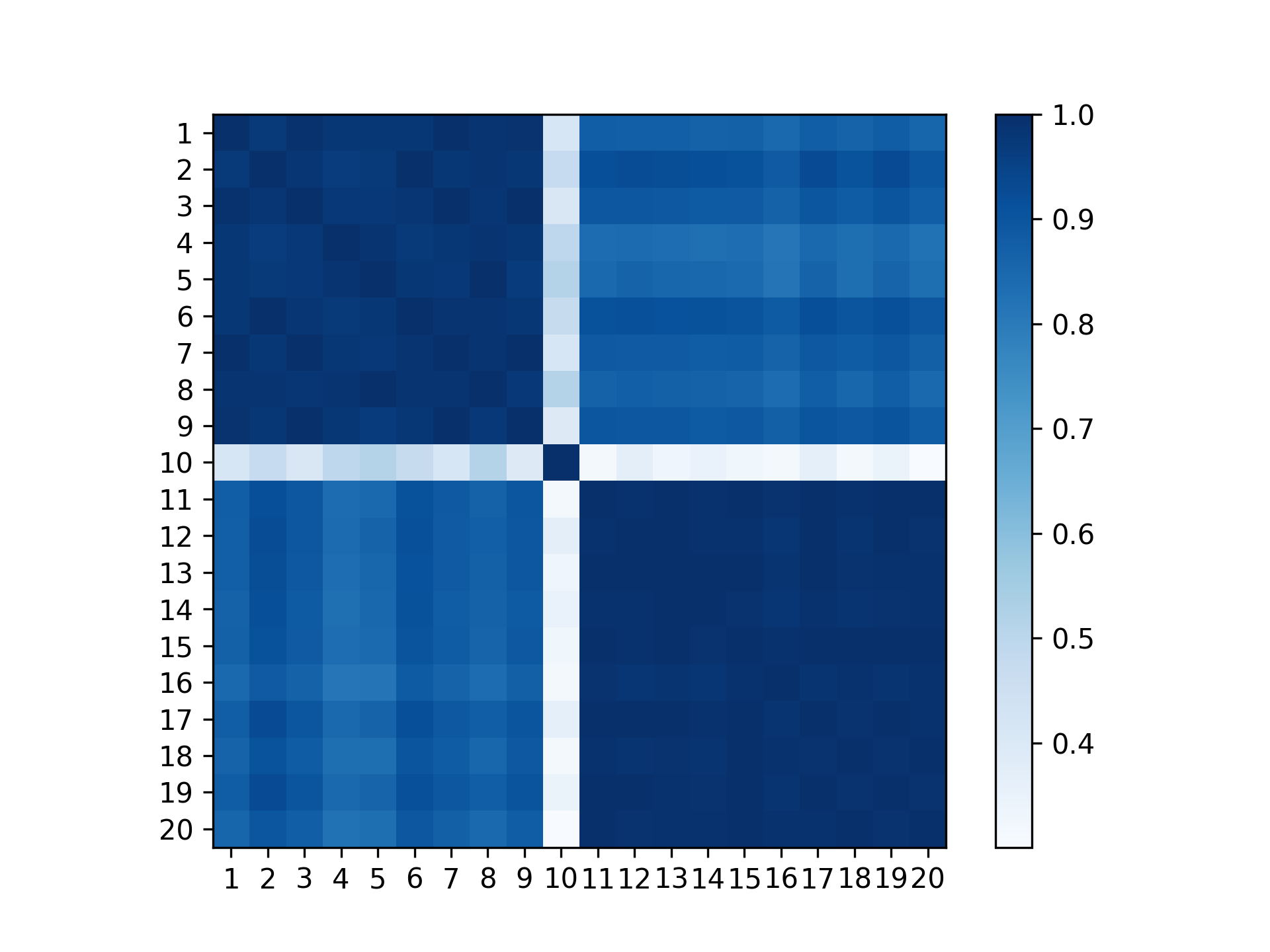}
    \caption{Pairwise Pearson's correlation between the (expected) posterior distribution of each event GP in the combined model. }
    \label{fig:corr_matrix}
\end{figure}

Rather than focus on prediction metrics here, we present how the correlation structure might inform monitoring procedures. %
One simple approach plots the correlation between the (expectation) of each $f_k$ posterior in the combined model, presented in \Cref{fig:corr_matrix}. %
Intuitively, there are two blocks within this confusion matrix: one associated with each train type. %
Additionally, the abnormal 350-event (index 9 within the upper left block) would clearly be identified as outlying: data which are not flagged by the decision tree heuristic (\Cref{fig:decision-tree}). %

\section{Concluding Remarks}\label{s:conc}
We have presented initial work towards modelling multilevel data structures, actively streaming from the (IB5) Staffordshire rail bridge. %
Firstly, a pre-existing bridge-weigh-in-motion system was used to classify train events into groups of a specific vehicle type -- here, we focus on commuter trains. %
For repeated crossings, the data were treated as longitudinal (or panel) data. %
A hierarchical (low-rank approximation) Gaussian process was then used to model aggregate train events, working towards a general representation of the strain response. %

Looking forward, it would be useful to parametrise the mean function, to learn a \textit{typical} representation for each \textit{train type}, alongside event-specific models. %
A basis function formulation using \textit{strain influence lines} will be investigated. %
An interpretable mean function should enable further insights regarding the variation of strain response, as well as the option for more expressive simulations. %

\section*{Acknowledgements}
The authors would like to thank Professor C.R.\ Middleton, Dr.\ F.\ Huseynov and P.R.A.\ Fiddler from the University of Cambridge, who led the latest phase of the instrumentation of the Norton Bridge, thanks to the support of the Centre for Digital Built Britain’s (CDBB). %
CDBB is a core partner of the Construction Innovation Hub, funded by UK Research and Innovation (UKRI) through the Industrial Strategy Challenge Fund (ISCF). %

LAB and MG acknowledge the support of the UK Engineering and Physical Sciences Research Council (EPSRC) through the ROSEHIPS project (Grant EP/W005816/1). %
AD is supported by Wave 1 of The UKRI Strategic Priorities Fund under the EPSRC Grant EP/T001569/1 and EPSRC Grant EP/W006022/1, particularly the \textit{Ecosystems of Digital Twins} theme within those grants \& The Alan Turing Institute. %

Finally, this research was possible thanks to the support of the 2022 Trimble Fund of The University of Cambridge.

\bibliographystyle{unsrtnatemph}
\bibliography{refs}

\end{document}